\documentclass[aps,pra,twocolumn,superscriptaddress]{revtex4-1}
\usepackage{amsmath,amsfonts,amssymb,bbm}
\usepackage{graphicx,subfigure}
\usepackage[colorlinks,urlcolor=black,linkcolor=blue,anchorcolor=blue,citecolor=blue]{hyperref}
\usepackage{mathtools}
\usepackage{multirow}
\usepackage{xcolor}
\usepackage{dcolumn}

\newcommand{\tr}[2][]{\textnormal{tr}#1{{\left\{#2\right\}}}}
\newcommand{\Eq}[2][Eq.~]{#1(\ref{eq:#2})}

\newcommand{\dket}[1]{{\vert{#1}\rangle\!\rangle}}
\newcommand{\dbra}[1]{{\langle\!\langle{#1}\vert}}
\newcommand{\map}[1]{{\mathcal{#1}}}

\newcommand{\rrangle}{\rangle\!\rangle}
\newcommand{\llangle}{\langle\!\langle}

\newcommand{\cC}{\mathcal{C}}

\newcommand{\cE}{\mathcal{E}}

\newcommand{\cG}{\mathcal{G}}

\newcommand{\cI}{\mathcal{I}}

\newcommand{\cO}{\mathcal{O}}

\newcommand{\cQ}{\mathcal{Q}}

\newcommand{\fg}{\mathfrak{g}}

\newcommand{\rhoin}{\rho_\mathrm{in}}
\newcommand{\gt}{\widetilde g}
\newcommand{\Mt}{\widetilde M}
\newcommand{\rhot}{\widetilde \rho}
\newcommand{\Pt}{\widetilde P}
\newcommand{\id}{\mathbbm{1}}
\newcommand{\eprep}[1]{e_{\mathrm{in}#1}}
\newcommand{\emeas}[1]{e_{\mathrm{ro},#1}}
\newcommand{\up}[1]{{(#1)}}
\newcommand{\upe}{\mathrm{e}}
\newcommand{\upi}{\mathrm{i}}
\newcommand{\upd}{\mathrm{d}}

\newcommand{\ve}{{\mathbf{e}}}
\newcommand{\vp}{{\mathbf{p}}}
\newcommand{\vC}{{\mathbf{C}}}
\newcommand{\vD}{{\mathbf{D}}}
\newcommand{\vV}{{\mathbf{V}}}
\newcommand{\vW}{{\mathbf{W}}}
\newcommand{\va}{{\boldsymbol{a}}}
\newcommand{\vb}{{\boldsymbol{b}}}
\newcommand{\valpha}{{\boldsymbol{\alpha}}}
\newcommand{\CNOT}{\mathrm{CNOT}}

\newcommand{\vigo}{{\tt ibmq{\textunderscore}vigo}}

\newcommand{\ourense}{{\tt ibmq{\textunderscore}ourense}}

\newcommand{\bogota}{{\tt ibmq{\textunderscore}bogota}}

\begin{document}

\title{Randomized linear gate set tomography}

\author{Yanwu \surname{Gu}}
\email{a0129460@u.nus.edu}
\affiliation{Yale-NUS College, Singapore}
\affiliation{Centre for Quantum Technologies, National University of Singapore, Singapore}

\author{Rajesh \surname{Mishra}}
\affiliation{Yale-NUS College, Singapore}
\affiliation{Centre for Quantum Technologies, National University of Singapore, Singapore}

\author{Berthold-Georg Englert}
\affiliation{Centre for Quantum Technologies, National University of Singapore, Singapore}
\affiliation{Department of Physics, National University of Singapore, Singapore}
\affiliation{MajuLab, CNRS-UCA-SU-NUS-NTU International Joint Research Unit, Singapore}

\author{Hui Khoon \surname{Ng}}
\email{huikhoon.ng@yale-nus.edu.sg}
\affiliation{Yale-NUS College, Singapore}
\affiliation{Centre for Quantum Technologies, National University of Singapore, Singapore}
\affiliation{MajuLab, CNRS-UCA-SU-NUS-NTU International Joint Research Unit, Singapore}

\date{\today}

\begin{abstract}
Characterizing the noise in the set of gate operations that form the building blocks of a quantum computational device is a necessity for assessing the quality of the device. Here, we introduce \emph{randomized linear gate set tomography}, an easy-to-implement gate set tomography procedure that combines the idea of state-preparation-and-measurement-error-free characterization of standard gate set tomography with no-design randomized tomographic circuits and computational ease brought about by an appropriate linear approximation. We demonstrate the performance of our scheme through simulated examples as well as experiments done on the IBM Quantum Experience Platform. In each case, we see that the performance of our procedure is comparable with that of standard gate-set tomography, while requiring no complicated tomographic circuit design and taking much less computational time in deducing the estimate of the noise parameters. This allows for straightforward on-the-fly characterization of the gate operations in an experiment.
\end{abstract}

\maketitle


With the recent advent of quantum computing devices, albeit small-scale ones only, there has been a lot of interest in understanding the nature of imperfections in the quantum gate operations, as a gauge of the quality of the computation as well as the hurdles to overcome in the quest for larger devices. Quantum process tomography \cite{QSETextbook} is the conventional method for characterizing the noise in gate operations. One prepares a specific set of quantum states as inputs to the gate of interest and measures a specific set of observables at the output. A comparison of the measurement results with the expected outputs from an ideal no-noise gate gives an estimate of the noise process of the actual gate operation. To do quantum process tomography well thus requires well-calibrated input preparation and measurement procedures, much better characterized than the noisy gate of interest. The current experimental reality, however, is such that the preparation and measurement procedures suffer from equally severe, and---more problematically---unstable, noise as the gates of interest, the so-called SPAM (for \underline{s}tate \underline{p}reparation \underline{a}nd \underline{m}easurement) errors. A separate pre-calibration, prior to the process tomography, is often not good enough to get rid of the effects of SPAM errors in the tomography results.

A more feasible approach is to characterize simultaneously the noise in the preparation and measurement procedures, alongside the noise in the gate operation itself. Earlier self-calibration ideas \cite{Mogilevtsev2009,Mogilevtsev2010,Branczyk2012,Quesada2013,Williams2017}, for state tomography but in principle applicable to process tomography, seek to deduce parameters of the SPAM procedures from the same tomographic data. More recently, a suite of methods under the heading of gate set tomography (GST) \cite{merkel2013self,blume2013robust,blume2017demonstration,greenbaum2015introduction} has emerged as a comprehensive approach to characterizing a ``gate set"---a specified set of gates that can be implemented in a quantum information processing device, for both computational tasks as well as the SPAM procedures---while avoiding the issue of SPAM errors. GST has been employed in various experiments; see, for example, Refs.~\cite{kim2015microwave,dehollain2016optimization,rol2017restless,chen2019detector,song2019quantum,rudinger2019probing}.

In short, GST handles the issue of SPAM errors by acknowledging the presence of noise in the SPAM procedures within the estimation itself. The task is no longer the characterization of a specific gate using other pre-calibrated gates in SPAM, but to characterize all the gates, those used in the computation as well as those in the SPAM procedures, from the same tomographic data, reminiscent of the early self-calibration work in state tomography. GST uses circuits---sequences of gates taken from the gate set of interest---with the structure of quantum process tomography: an initial segment of gates corresponding to the preparation, from an initial default state, of a tomographic input state; a middle segment comprising the gate(s) of interest; a final segment containing the gates that, together with a final default measurement, implement the measurement of a particular tomographic observable. In the actual GST noise reconstruction procedure, no difference is made between the gates from the three segments---all gates are considered as noisy, and the noise parameters of the entire gate set are the targets of the reconstruction---but this underlying structure is the basis of the GST circuit design.

In this work, we introduce a simpler alternative to GST, offering speed of reconstruction (seconds compared to half an hour for GST in a two-qubit case) based on a linear noise regime, easy circuit design (randomized compared to the carefully designed structured GST circuits), and applicability to tomographically incomplete gate sets, not naturally accessible within the GST framework, while yielding results quite comparable to those from GST. GST, in its entirety, can give a more accurate reconstruction of the noise, but if one is looking for an on-the-fly and easy-to-implement approximate characterization of the current noise in a device, our \emph{randomized linear gate set tomography} (RL-GST) offers a quick and simple solution. Potentially, the results from our RL-GST approach can also be used as the starting points for the optimization procedure in the usual GST, for a more refined noise estimate. The approximation of a linear noise regime, as we will see, restricts RL-GST to circuit lengths short compared with the inverse of the noise-strength, but such  a requirement is natural anyway for accurate calculations by the quantum devices.

Below, we first describe our RL-GST procedure, explaining the linear noise regime and the choice of randomized circuits. We follow that with examples, simulated ones as well as experiments on the IBM Quantum Experience Platform \cite{ibmQX}, to illustrate the performance of RL-GST, in comparison with that of standard GST.

\bigskip
\bigskip
\noindent\textbf{TOMOGRAPHY IN THE LINEAR REGIME}\\[1ex]
\textbf{Preliminaries. }Consider a quantum computing platform capable, ideally, of performing operations from a gate set $\fg\equiv \{g_\gamma\}_{\gamma=1}^\Gamma$, preparing (initialization) a default pure input state $|\psi_\mathrm{in}\rangle$, and measuring (readout) a default observable $M\equiv\sum_\mu\lambda_\mu P_\mu$, with $\lambda_\mu$s the eigenvalues of $M$, and $P_\mu$s the corresponding eigenoperators, which are projectors not necessarily of rank 1. Here, we assume a discrete set of $\Gamma$ gates for an $n$-qubit system. Each $g_\gamma$ is a unitary map acting on an operator on the $n$-qubit Hilbert space as $g_\gamma(\cdot)=U_\gamma(\cdot) U_\gamma^\dagger$, where $U_\gamma$ is a unitary operator. 

In a real platform, every component will be noisy. We denote the noisy version of each component by a tilde: $\gt_\gamma$ is the noisy implementation of the ideal gate $g_\gamma$, $\rhot_\mathrm{in}$ is the noisy initialization of the ideal input density operator $\rho_\mathrm{in}\equiv |\psi_\mathrm{in}\rangle\langle\psi_\mathrm{in}|$, and $\Mt=\sum_\mu\lambda_\mu\Pt_\mu$ is the noisy version of the ideal readout $M$. While $g_\gamma$ is a unitary map, $\gt_\gamma$ is assumed to be a general---not necessarily unitary---completely positive (CP) and trace-preserving (TP) linear map. We write the noisy gate as $\gt_\gamma=(\id+e_\gamma)g_\gamma$, so that $\id+e_\gamma\equiv \gt_\gamma g_\gamma^{-1}$ is the CPTP noise map for the gate $g_\gamma$, and $e_\gamma$ is its deviation from the identity map $\id$. Note that $e_\gamma$ is not TP; in fact, $\tr{e_\gamma(\cdot)}=0$ for any input. Neither is $e_\gamma$ necessarily CP; only $\id+e_\gamma$ is CP. Similarly, we assume that $\rhot_\mathrm{in}=(\id+\eprep{}) (\rho_\mathrm{in})$ and $\Pt_\mu=(\id+\emeas{\mu})^\dagger(P_\mu)$, for CPTP $\id+\eprep{}$ and $\id+\emeas{\mu}$. $\rhot_\mathrm{in}=(\id+\eprep{}) (\rho_\mathrm{in})$ can be read as having the ideal input $\rho_\mathrm{in}$ followed by the application of the noise map $\id+\eprep{}$; $\Pt_\mu=(\id+\emeas{\mu})^\dagger(P_\mu)$ should be, because of the adjoint, read instead as the application of the noise map $\id+\emeas{\mu}$, following by the ideal projector $P_\mu$. Note that the set of $\emeas{\mu}$s are not all independent: We must have $\sum_\mu\widetilde P_\mu=\id$, for a valid measurement. For the case of weak noise, necessary for a useful quantum computing platform, we expect $\Vert e_\gamma\Vert,\Vert \eprep{}\Vert,\Vert\emeas{\mu}\Vert\ll \Vert\id\Vert$, for some map norm $\Vert\cdot\Vert$. It is convenient to assume $\Vert \id\Vert =1$, and that $\Vert e_\gamma\Vert,\Vert \eprep{}\Vert,\Vert\emeas{\mu}\Vert\lesssim \epsilon$ for some positive $\epsilon\ll 1$, for all $\gamma$; $\epsilon$ will be the book-keeping parameter for our perturbative analysis below. 

In the operation of the quantum computer, following what is sometimes referred to as the standard circuit model, the default input state is prepared, and a circuit $\cC$---a sequence of gates from the gate set $\fg$---is performed on that input state, and the default observable $M$ is measured. $\cC$ contains not just the gate sequence that carries out the desired computation on some input state. It includes, as the starting sequence, the gates that prepare the default input state $|\psi_\mathrm{in}\rangle$ into the actual desired input state, and, as the ending sequence, the gates that, together with the readout $M$, implement a desired measurement on the output state:
\begin{equation}
\cC=\underbrace{\cG_L\cG_{L-1}\cdots}_{\mathrm{measurement}}\underbrace{\cdots \cG_{k+1}\cG_k\cG_{k-1}\cdots}_{\mathrm{computation}}\!\!\underbrace{\cdots\cG_2\cG_1}_{\mathrm{preparation}}\!, ~~\cG_k\!\in\!\fg~\forall k.
\end{equation}
In reality, we have the noisy version of the ideal circuit $\cC$, written as $\widetilde\cC=\widetilde\cG_L\cdots\widetilde\cG_1$, where $\widetilde\cG_k$ is the noisy implementation of the ideal gate $\cG_k$. As for the noisy version of the gate set elements, we write $\widetilde\cG_k=(\id+\cE_k)\cG_k$, where $\id+\cE_k$ is assumed to be CPTP. We assume a sequence-independent noise model, i.e., $\cE_k=e_\gamma$ whenever $\cG_k=g_\gamma$.

We introduce some convenient notation before proceeding further. We can ``vectorize" the operators on the $n$-qubit Hilbert space, and write $|O\rrangle$ as the vectorized operator $O$, now considered as a vector in the vector space of operators. The adjoint vector $\llangle O|$ is defined such that the Hilbert--Schmidt inner product between two operators $O_1$ and $O_2$ can be written as $\tr{O_1^\dagger O_2}=\llangle O_1|O_2\rrangle$. We write the action of a map $\cG$ on an operator $O$ as $\cG(O)=\cG|O\rrangle$, where we have used, for conciseness, the same symbol for the map $\cG(\cdot)$ on the left, and the (super)operator (matrix) $\cG$ representing the map on the right, acting on the vector (column) $|O\rrangle$. Then, we can write the entire quantum computation, from preparing the input state $\rho_\mathrm{in}$, to implementing the circuit $\cC$, to getting the measurement outcome $\mu$, as $\llangle P_\mu|\cC|\rho_\mathrm{in}\rrangle$. Its noisy version is, correspondingly, $\llangle \Pt_\mu|\widetilde\cC|\rhot_\mathrm{in}\rrangle$. We will also use the shorthand $\cG_{m:l}$, for positive integers $m$ and $l$, to denote the gate sequence $\cG_m\cG_{m-1}\cdots\cG_{l+1}\cG_l$ if $m\geq l$, and the $\id$ map if $m<l$; for example, $\cC=\cG_{L:1}$, $\cG_{1:1}=\cG_1$, and $\cG_{1:2}=\id$. In addition, we write ${}_\mu\!\llangle \,\cdot\, \rrangle$ as a shorthand for $\llangle P_\mu|\,\cdot\,|\rhoin\rrangle$.

\medskip
\noindent \textbf{Linear noise regime. }We are now ready to define the linear regime for the noise. The output of the quantum computer running circuit $\cC$ are the outcome probabilities $\{p_\mu\equiv \llangle P_\mu|\cC|\rho_\mathrm{in}\rrangle={}_\mu\!\llangle\cC\rrangle\}$. We write the noisy version $\widetilde p_\mu$ as an ``error expansion", organized in powers of $\epsilon$,
\begin{align}\label{eq:lin}
\widetilde p_\mu&=\llangle \Pt_\mu|\widetilde\cC|\rhot_\mathrm{in}\rrangle\nonumber\\
&={}_\mu\!\llangle \cC\rrangle\hspace{5cm}\textrm{[$\cO(1)$ terms]}\nonumber\\
&+{}_\mu\!\llangle \cC\eprep{} +\sum_{k=1}^L \cG_{L:k+1}\cE_k\cG_{k:1}+\emeas{\mu}\cC\rrangle\quad\textrm{[$\cO(\epsilon)$ terms]}\nonumber\\
& +{}_\mu\!\llangle \cG_{L:2}\cE_1\cG_1\eprep{}+\cG_{L:3}\cE_{2}\cG_2\cE_1\cG_1+\!\ldots\rrangle~\textrm{[$\cO(\epsilon^2)$ terms]}\nonumber\\
& +\ldots.
\end{align}
The linear noise regime is defined as the linear truncation of this error expansion, keeping up to, and including, the $\cO(\epsilon)$ terms, and dropping $\cO(\epsilon^2)$ and higher-order terms. This corresponds to at most one insertion of the ``error" $e_\gamma, \emeas{\mu}$, or $\eprep{}$ in each term in the error expansion. This linear regime is a good approximation only if 
\begin{equation}
L\epsilon\ll1,
\end{equation}
so that the neglected higher-order terms give small corrections.

Our goal is to estimate the errors $e_\gamma, \emeas{\mu}$, and $\eprep{}$, relevant for a quantum computing platform with its gate set and its initialization and readout procedures, well enough for making predictions of the outcomes $\{p_\mu\}$ of circuits run on the platform. We do this by running a set of tomographic circuits $\{\cC^\up{i}\}$ on the quantum computing platform and recording the outcome data. From the collected data, our adoption of the linear noise regime permits a straightforward linear-inversion procedure (see below) for obtaining estimates of the error maps $e_\gamma,\emeas{\mu}$, and $\eprep{}$. The linear noise regime captures the dominant sources of noise in the platform, and the errors estimated within this regime suffice to make accurate predictions of the outcomes of circuits with length $\lesssim 1/\epsilon$.

\medskip
\noindent \textbf{Linear inversion.}
Let us first rewrite Eq.~\eqref{eq:lin} in the linear regime, for circuit $\cC^\up{i}$, as
\begin{align}\label{eq:lin2}
&\quad~ \widetilde p^\up{i}_\mu-p^\up{i}_\mu\\
&\simeq {}_\mu\!\llangle \cC^\up{i}\eprep{}+\sum_{k=1}^L \cG^\up{i}_{L:k+1}\cE_k^\up{i}\cG^\up{i}_{k:1}+\emeas{\mu}\cC^\up{i}\rrangle\nonumber\\
&={}_\mu\!\llangle \cC^\up{i}\eprep{}+\sum_{\gamma=1}^\Gamma\sum_{k\in\mathrm{Pos}^\up{i}_\gamma}\!\!\! \cG^\up{i}_{L:k+1}e_\gamma g_\gamma\cG^\up{i}_{k-1:1}+\emeas{\mu}\cC^\up{i} \rrangle,\nonumber
\end{align}
where $\mathrm{Pos}_\gamma^\up{i}$ denotes the subset of indices $k$ corresponding to the positions in the circuit $\cC^\up{i}=\cG^\up{i}_L\cG^\up{i}_{L-1}\ldots \cG^\up{i}_1=\cG^\up{i}_{L:1}$ where $\cG_k^\up{i}=g_\gamma$. For concreteness, we introduce an orthonormal basis $\{|B_a\rrangle\}_{a=1}^{d^2}$ ($d$ is the dimension of the Hilbert space), and write the gate errors as $e_\gamma=\sum_{ab}e_{\gamma;ab}|B_a\rrangle\llangle B_b|$, for complex matrix elements $e_{\gamma;ab}\equiv \llangle B_a|e_\gamma|B_b\rrangle$. In addition, we write the noisy initialization as $\eprep{}|\rhoin\rrangle = \sum_a\eprep{;a}|B_a\rrangle$, for complex scalars $\eprep{;a}\equiv \llangle B_a|\eprep{}|\rhoin\rrangle$; the noisy readout is, similarly, $\llangle P_\mu|\emeas{\mu}=\sum_a\emeas{\mu;a}\llangle B_a|$ for $\emeas{\mu;a}\equiv \llangle P_\mu|\emeas{\mu}|B_a\rrangle$. The complex parameters $e_{\gamma;ab}$, $\eprep{;a}$, and $\emeas{\mu;a}$ fully describe the noise in the platform, and are the quantities we want to estimate. 

In terms of these noise parameters, inserting the identity $\sum_a|B_a\rrangle\llangle B_a|$ at appropriate places, Eq.~\eqref{eq:lin2} can be rewritten as
\begin{align}\label{eq:lin3}
&\quad~ \widetilde p^\up{i}_\mu-p^\up{i}_\mu\simeq c_{\mu;a}^\up{i}\eprep{;a}+c^\up{i}_{\mathrm{in},\mu\gamma;ab}e_{\gamma;ab}+c^\up{i}_{\mathrm{in};a}\emeas{\mu;a},
\end{align}
with the understanding that repeated indices are summed. Here, the various $c^\up{i}$ coefficients are complex scalars determined solely by the ideal circuit $\cC^\up{i}$ and our basis choice: $c^\up{i}_{\mathrm{in};a}\equiv \llangle B_a|\cC^\up{i}|\rhoin\rrangle$, $c^\up{i}_{\mathrm{in},\mu\gamma;ab}\equiv \sum_{k\in\mathrm{Pos}_\gamma^\up{i}}\llangle P_\mu|\cG^\up{i}_{L:k+1}|B_a\rrangle\llangle B_b|g_\gamma\cG^\up{i}_{k-1:1}|\rhoin\rrangle$, and $c_{\mu;a}^\up{i}\equiv \llangle P_\mu|C^\up{i}|B_a\rrangle$.
Each circuit $\cC^\up{i}$ gives a set of equations, Eq.~\eqref{eq:lin3} for different $\mu$s, for the noise parameters $e$, with $\widetilde p^\up{i}$ estimated from the experimental data, and $p^\up{i}$ and $c^\up{i}$ coefficients computed from the ideal $\cC^\up{i}$ and chosen basis. 
Stacking up the equations, for some choice of ordering of $\mu$ and $i$, we can write down a system of linear equations for $e$,
\begin{equation}\label{eq:linsys}
\widetilde{\vp}-\vp=\vC\ve,
\end{equation}
where $\widetilde{\vp}$($\vp$) is the column of $\widetilde p_\mu^{(i)}$($p_\mu^\up{i}$), $\ve$ is the column of noise parameters, and $\vC$ is a (generally non-square) matrix of the $c$ coefficients, with each row comprising the $c$s for the same $\mu$ and $i$. We refer to $\vC$ as the \emph{design matrix}, to emphasize its dependence only on our choice of circuits and basis, not on the noise maps of the system. The noise parameters are then estimated by solving Eq.~\eqref{eq:linsys} for $\ve$. We do this by linear inversion, which essentially amounts to writing $\ve=\vC^{-1}(\widetilde{\vp}-\vp)$, provided we can make sense of the inverse of $\vC$. This is what we discuss next.

A remark, before we discuss the design matrix further: The linear noise regime is a natural approximation in the weak-noise limit, the expected scenario for any useful quantum computing platform, in many situations, including randomized benchmarking. In particular, it was used previously, also with an accompanying linear-inversion procedure, in the pre-cursor, Ref.~\cite{merkel2013self}, to the current formulation of GST. There, however, the characterization circuits all follow the SPAM-gates structure of standard GST, with a single application of the gate-of-interest in between. Here, our characterization circuits, as we will see, are instead randomized ones. Our linear gate set tomography procedure is also different from that in Ref.~\cite{blume2013robust} carrying the same name: In Ref.~\cite{blume2013robust}, the ``linear" tag refers solely to the linear inversion employed in the estimation, with the linear structure achieved with special choices of GST-type circuits, without the use of the linear-noise approximation as we do here.

\medskip
\noindent\textbf{Dependence relations. }
The design matrix $\vC$ can be non-invertible for a few reasons. First, $\vC$ is a non-square matrix, as we can have more equations, from the different choices of $\cC^\up{i}$ and different $\mu$ values, than there are $\ve$ parameters. This in itself is not a problem: We only need to include in $\vC$ a linearly independent set of equations found, for example, by a row-reduction procedure on the original $\vC$. That is not all, however. That $\id +e_\gamma$, $\id+\eprep{}$, and $\id+\emeas{\mu}$ are all CP (and hence hermicity-preserving) and TP maps translate to equality constraints on the $e$ parameters, so that not all $e$ parameters are independent in the first place. Furthermore, dependence relations between the columns of the design matrix---leading to nonunique solutions for $\ve$---arise because of what is sometimes known as ``gauge freedom", coupled with our linear regime approximation.

We can easily accommodate the CPTP conditions for $n$-qubit situations by working in the Pauli operator basis: the set of Pauli operators together with the identity operator, $\{\frac{1}{\sqrt 2}\sigma_a\}_{a=0}^3=\frac{1}{\sqrt 2}\{\id, \sigma_ 1=\sigma_x,\sigma_2=\sigma_y, \sigma_3=\sigma_z\}$ for a single qubit, and the tensor products $\{\frac{1}{(\sqrt 2)^n}\sigma_{a_1}\otimes\sigma_{a_2}\ldots\sigma_{a_n}\}$, for $a_1, a_2,\ldots, a_n=0,1,2,3$ for $n$ qubits. We write an element of the $n$-qubit Pauli basis as $\frac{1}{(\sqrt 2)^n}\sigma_{\va}$ for $\va\equiv (a_1,a_2,\ldots, a_n)$. Note that $\sigma_{\boldsymbol{0}}$ is always the element proportional to the identity operator, and is the only basis element with nonvanishing trace.

Now, a TP map cannot take a traceless Pauli operator to the identity with nonvanishing trace. Thus, in the Pauli basis, the TP condition requires $e_{\gamma;\boldsymbol{0}\va}=0$ for all $\va$ (recall: $\id+e_\gamma$, not $e_\gamma$ itself, is TP). The CP condition is first a hermicity-preserving condition, which in the Pauli basis, translates into the statement that $e_{\gamma;\va\vb}$ are all real, not complex, numbers. That it is CP adds only inequality constraints that do not reduce the number of independent noise parameters. Thus, each gate error $e_\gamma$ is determined by no more than $d^2(d^2-1)$ real parameters.
For the initialization and readout error maps, we are concerned only with those matrix elements picked out by $|\rhoin\rrangle$ and $\llangle P_\mu|$, namely, $\eprep{;\va}=\llangle\sigma_\va |\eprep{}|\rhoin\rrangle$ and $\emeas{\mu;\va}=\llangle P_\mu|\emeas{\mu}|\sigma_\va\rrangle$. That $\id+\eprep{}$ is TP means $\eprep{;\boldsymbol{0}}=\llangle\sigma_{\boldsymbol{0}} |\eprep{}|\rhoin\rrangle=\frac{1}{\sqrt d}\tr{ \eprep{}(\rhoin)}=0$, for any $\rhoin$. We are thus left with $d^2-1$ real (since $\id + \eprep{}$ is CP) parameters to specify the initialization error. For the readout, the TP condition does not impose any additional constraints, and the CP condition translates into a total of $d^2$ real parameters to specify each $\emeas{\mu}$, subject to the additional overall constraint $\sum_\mu\widetilde P_\mu=\id$.

In addition, we have to consider dependences amongst the columns of the design matrix due to gauge freedom. Gauge freedom in this context, following the terminology in earlier papers \cite{Proctor2017,merkel2013self,blume2013robust,blume2017demonstration}, refers to the freedom to attribute the observed noise to the gate imperfections versus the initialization and readout (or SPAM) errors. This stems from our ability to probe the physical device only by sending in a single default input state and getting the readout values for a single default observable; one cannot access the gate-only noise without involving also the initialization and readout errors. In our linear noise regime, this gauge freedom is expressible as the fact that we can only access the values of the sum ${}_\mu\!\llangle \cC\eprep{}\rrangle +\sum_{k=1}^L {}_\mu\!\llangle\cG_{L:k+1}\cE_k\cG_{k:1}\rrangle+{}_\mu\!\llangle\emeas{\mu}\cC\rrangle$, not the values of the individual terms.
Assuming that we have removed the redundancy in $\vC$ arising from the CPTP constraints as well as those from linear dependences in the circuit choices, the presence of this gauge freedom can be phrased, in our linear regime, as the nontrivial null space of $\vC$. A nontrivial null space translates into the fact that any two solutions $\ve$ and $\ve'$ of Eq.~\eqref{eq:linsys} differ by a vector $\valpha$ in the null space. The null space of $\vC$ can easily be found numerically by performing a singular-value decomposition on $\vC$. 

The addition of a null-space vector $\valpha$ to $\ve$ to obtain $\ve'$ can be expressed as a transformation on the original $e$ noise maps. A particular transformation that generates a gauge freedom in our linear noise regime is
\begin{align}
e_\gamma&\rightarrow e_\gamma + \cQ-g_\gamma \cQ g_\gamma^\dagger,\nonumber\\
\label{eq:gaugeTransf}\eprep{}|\rhoin\rrangle&\rightarrow \eprep{}|\rhoin\rrangle+\cQ|\rhoin\rrangle,\\
\textrm{and}\quad \llangle P_\mu|\emeas{\mu}&\rightarrow \llangle P_\mu|\emeas{\mu}-\llangle P_\mu|\cQ,\nonumber
\end{align}
for any $\gamma$-independent linear map $\cQ$. This is a gauge transformation in that it preserves the value of ${}_\mu\!\llangle \cC\eprep{}\rrangle +\sum_{k=1}^L {}_\mu\!\llangle\cG_{L:k+1}\cE_k\cG_{k:1}\rrangle+{}_\mu\!\llangle\emeas{\mu}\cC\rrangle$ for all $\mu$, a statement that can be checked explicitly. Every choice of $\cQ$ leads to an equivalent set of noise maps, equivalent in terms of their effects on the circuit output within the linear noise regime. While we cannot, at the moment, show that the form of Eq.~\eqref{eq:gaugeTransf} fully captures the gauge freedom allowed within our linear noise regime, the full null spaces in our illustrative examples below can all be described in this manner; see Supplementary information.

One should compare our gauge transformation Eq.~\eqref{eq:gaugeTransf} to the analogous transformation found in the GST papers. In GST, the appropriate gauge transformation is given by \cite{merkel2013self,blume2013robust,blume2017demonstration}
\begin{align}
\widetilde g_\gamma &\rightarrow M \widetilde g_\gamma M^{-1}, \nonumber\\
\label{eq:gaugeTransfGST}|\widetilde{\rho}_\mathrm{in}\rrangle &\rightarrow M|\widetilde{\rho}_\mathrm{in}\rrangle,\\
\textrm{and}\quad\llangle{\widetilde{P}_\mu}| &\rightarrow \llangle\widetilde{P}_\mu|M^{-1},\nonumber
\end{align}
for any invertible $M$. To compare with Eq.~\eqref{eq:gaugeTransf}, we first note that the first line of Eq.~\eqref{eq:gaugeTransf} can be re-written in terms of $\widetilde g_\gamma$ as $\widetilde g_\gamma\rightarrow \widetilde g_\gamma +\cQ g_\gamma - g_\gamma \cQ$. Now, we suppose our $\cQ$ can be regarded as a small quantity, and that $M=\upe^{\cQ}\simeq \id+\cQ$. Then, the first line of Eq.~\eqref{eq:gaugeTransfGST} can be written as $\widetilde g_\gamma\rightarrow M \widetilde g_\gamma M^{-1}\simeq (\id+\cQ)(g_\gamma+e_\gamma g_\gamma)(\id-\cQ)\simeq\widetilde g_\gamma +\cQ g_\gamma-g_\gamma\cQ$, dropping terms second-order in small quantities ($\cQ$ and $e_\gamma$), yielding our transformation Eq.~\eqref{eq:gaugeTransf} for $\widetilde g_\gamma$. Similar relations for the transformations on $|\widetilde \rho_\mathrm{in}\rrangle$ and $\llangle\widetilde P_\mu|$ hold as well.

\medskip
\noindent \textbf{The RL-GST procedure.} Finally, the procedure for our RL-GST approach is as follows:
\begin{enumerate}
\item Choose a set of $N_c$ tomographic circuits $\{\cC^\up{i}\}$, with $N_c$ larger than the number of noise parameters. Each circuit $\cC^{(i)}$ is built from a uniform-random sequence of gates from the gate set $\fg$.
\item Compute $\vp$ from the circuits. In addition, perform the circuits on the actual device, and obtain an estimate of $\widetilde{\vp}$ from the gathered data.
\item Calculate the design matrix $\vC$ from the circuit set, for a choice of operator basis. We can elect to use the Pauli basis, for ease in incorporating the CPTP conditions on our noise parameters.
\item Perform singular-value decomposition of $\vC=\vV\vD\vW^\dagger$, where $\vD$ is the diagonal (generally non-square) matrix of singular values $\lambda_j$s, and $\vV$ and $\vW$ are unitary matrices. The columns of $\vW$ corresponding to the zero singular values span the null space of $\vC$.
\item We construct the (pseudo)inverse of $\vC$ as $\vC^{-1}=\vW\vD^{-1}\vV^\dagger$ where $\vD^{-1}\equiv \textrm{diag}\{\lambda'_1,\lambda'_2,\ldots\}$ with $\lambda'\equiv 1/\lambda_j$ if $\lambda_j\neq 0$, and $0$ otherwise.
\item Then, an estimate $\widehat\ve$ for the noise parameters $\ve$ is obtained by direct inversion,
\begin{equation}
\widehat\ve=\vC^{-1}(\widetilde{\vp}-\vp).
\end{equation}
\end{enumerate}
Here, as we are looking for a quick estimate of the noise, and in the tradition of linear-inversion estimators, we do not impose any CP constraints on our estimator. The $\id+\widehat e_\gamma$ noise maps obtained from $\widehat\ve$ will thus generally not be CPTP. If the CPTP nature is crucial, one can further apply a standard procedure to project the obtained estimator to the nearest CPTP map, for example, by the procedure of Ref.~\cite{Knee2018}. 

As an added step, one could perform a singular value decomposition of the design matrix before carrying out the tomographic circuits. This gives the number of zero singular values in the design matrix, yielding useful information about our choice of tomographic circuits. In particular, if we find that to be larger than $(d^2-1)d^2$, the number of singular values caused by Eq.~\eqref{eq:gaugeTransf} (see Supplemental information), then we should figure out if these extra zero singular values arise from an unlucky sampling of tomographic circuits or from some other kinds of gauge freedom beyond Eq.~\eqref{eq:gaugeTransf} by sampling more circuits or adding some short circuits (see examples below, where we add ``null" circuits).

Note that each circuit $\cC^{(i)}$ should have length $L^{(i)}\ll 1/\epsilon$, for the linear regime to hold. $\epsilon$ can be estimated, a priori, using quick partial characterization methods like randomized benchmarking. $L^{(i)}$ should also generally not be too short---recalling the structure of quantum process tomography, where one should use a variety of input states and measure a variety of observables on the output, the circuit $\cC^{(i)}$ should have enough gates to give enough variety to sense all aspects of the error maps. As we will see in our illustrative examples below, the accuracy of the RL-GST procedure is not sensitive to the specific choice of $L^{(i)}$s, and in fact, one can use a variety of values for $L^{(i)}$ within the same circuit set. Furthermore, the circuit set can be pre-assessed, before doing the experiments, for its effectiveness in sensing the noise parameters: A good choice should have good linear independence between the rows of $\cC$, a notion that can be quantified in many ways. We have observed, however, that our randomized gate choice generally yield circuit sets with good linear independence, and have not found it necessary to do such a pre-assessment in our examples.

Armed with $\widehat\ve$, we can now make predictions from the device implementations of further ``test" circuits, and expect accurate estimates of the outputs of the noisy device, namely, the quantities $p_\mu(\cC)\equiv \llangle \widetilde P_\mu|\widetilde\cC|\widetilde \rho_\mathrm{in}\rrangle$ for test circuit $\cC$, as long as the length of the test circuit satisfies $L\epsilon\ll 1$. In our examples below, we will measure the prediction error using the statistical distance,
\begin{equation}\label{eq:statDist}
\Delta \equiv \frac{1}{2}\sum_\mu\vert \widehat{p}_\mu-p_\mu\vert,
\end{equation}
where $\widehat p_\mu$ is computed from our estimate $\widehat\ve$, while $p_\mu$ is the true value.

Because of the gauge freedom, we cannot say that our estimate $\widehat\ve$, one representative choice amongst those that differ by a null-space vector, gives the true values of the actual noise parameters. Thus, $\widehat\ve$ cannot be used to directly infer quantities like the strength of the noise, or the true noise maps, beyond this sandwich structure of $\llangle P_\mu|\cdot|\rho_\mathrm{in}\rrangle$. However, if the null-space equivalence can be fully captured by the gauge transformation Eq.~\eqref{eq:gaugeTransf} (as is the case in our examples below), then we can use $\widehat\ve$ to estimate a common measure for judging the quality of gates: the average gate set infidelity (AGsI), defined as
\begin{align}
\label{agsi}\mathrm{AGsI}(\fg)&\equiv\frac{1}{\Gamma}\sum_{\gamma=1}^\Gamma[1-\overline F(g_\gamma,\widetilde g_\gamma)],
\end{align}
with
\begin{align}
\overline F(g,g')&\equiv \int\upd\psi\langle\psi|(g^\dagger \circ g')(\psi)|\psi\rangle=\frac{\tr{g^\dagger g'}+d}{d(d+1)},
\end{align}
for $\upd\psi$ taken as the Haar measure. The final expression for $\overline F$ is a Haar-average formula familiar in many contexts (e.g., see Refs.~\cite{horodecki1999general,nielsen2002simple}). $\overline F(g_\gamma,\widetilde g_\gamma)$ is invariant under the gauge transformation Eq.~\eqref{eq:gaugeTransf}, i.e., $\tr{g_\gamma^\dagger (\id +e_\gamma)g_\gamma}=\tr{g_\gamma^\dagger (\id +e'_\gamma)g_\gamma}$ for $e'_\gamma = e_\gamma+\cQ-g_\gamma\cQ g_\gamma^\dagger$, since $\cQ-g_\gamma\cQ g_\gamma^\dagger$ is traceless. In our examples below, we thus use, apart from the prediction errors, the AGsI as another figure-of-merit for judging the performance of our RL-GST procedure. Note that the AGsI is \emph{not} invariant under the GST gauge transformation Eq.~\eqref{eq:gaugeTransfGST}, as previously noted in Ref.~\cite{Proctor2017}.

\bigskip
\bigskip
\noindent\textbf{RESULTS}\\[1ex]
To illustrate our method, we carry out simulated experiments for single- and two-qubit gate sets, as well as use our RL-GST procedure to characterize three 5-qubit IBM quantum computing devices available for free access. We compare the performance of our method against that of standard GST. For implementing the GST protocol, we make use of the Python package `pygsti' from Ref.~\cite{osti_1543289}.

\bigskip
\noindent\textbf{Single-qubit simulations.} 
To assess the performance of our RL-GST procedure, it is important to begin with a situation where we know the true noise in the gates. We simulate noisy single-qubit gate operations by attaching to an ideal gate a noise map of the form $e(\ve)\equiv \textrm{AD}(\upe_0)\circ \textrm{Pauli}(\upe_1,\upe_2,\upe_3)\circ R_x(\upe_4)\circ R_y(\upe_5)\circ R_z(\upe_6)$. Here, $\textrm{AD}(\upe_0)$ is the amplitude-damping channel $\textrm{AD}(\upe_0)(\cdot)=(|0\rangle\langle 0|+\sqrt{1-\upe_0}|1\rangle\langle 1|)(\cdot)(|0\rangle\langle 0|+\sqrt{1-\upe_0}|1\rangle\langle 1|)+\upe_0|0\rangle\langle 1|(\cdot)|1\rangle\langle 0|$, $\textrm{Pauli}(\upe_1,\upe_2,\upe_3)$ is the Pauli channel $\textrm{Pauli}(\upe_1,\upe_2,\upe_3)(\cdot)=[1-(\upe_1+\upe_2+\upe_3)](\cdot)+\sum_{a=1^3}\upe_a\sigma_a(\cdot)\sigma_a$, and $R_{X(Y,Z)}(\upe_{4(5,6)})$ is the rotation about the $x(y,z)$-axis of the Bloch sphere by angle $\upe_{4(5,6)}$. For each gate in the gate set, the noise parameters $\ve$ are randomly chosen, but they remain fixed for the entire tomography procedure. In addition, we set $\widetilde\rho_\mathrm{in}\equiv \widetilde\rho_\mathrm{in}^{(1)}=|0\rangle\langle 0|+\frac{a}{\sqrt 2} (X+Y-Z)$, and $\widetilde P_0$ is taken to be identical to $\widetilde\rho_\mathrm{in}$. The $a$ is set as $0.01$ in our examples below. Data for the tomography procedure are generated using $N_s=8192$ shots (or repetitions) per circuit.

\medskip
\noindent\textit{Incomplete gate set.} We first discuss the case of a tomographically incomplete---or just incomplete for short---gate set, by which we mean a gate set unable to produce a tomographically complete set of states starting with the default $|0\rangle$ input. A \emph{computationally} complete (or universal) gate set is always tomographically complete, but a tomographically complete gate set need not be computationally complete. An example of the latter is the single-qubit gate set $\{\id,X_{\pi/2}, Y_{\pi/2}\}$, where $X_{\pi/2}(Y_{\pi/2})$ is a $\pi/2$ rotation about the $x(y)$-axis of the Bloch sphere. An incomplete gate set means we are unable to sense all aspects of the gate noise (up to the gauge transformation discussed earlier), but we seek only to predict the outcomes of circuits built from that same gate set, for which the inaccessible noise parameters play no role.

An often-encountered incomplete gate set is the set of single-qubit Pauli gates $\{X\equiv\sigma_x, Y\equiv\sigma_y, Z\equiv\sigma_z\}$. For each circuit length $L^{(i)}=8,16,32,64,$ and $128$, we generate uniform-randomly, from the Pauli gate set, a set of $N_c=150$ circuits $\{C^{(i)}\}$. Duplicate circuits are discarded, and a null circuit, i.e., one with $L^{(i)}=0$, is always added as we found it prevents accidental generation of an extra zero singular value in the design matrix. We obtain the estimate $\widehat\ve$ for the noise parameters as described in the previous section, and use it to make predictions for a further 1000 test circuits, chosen uniform-randomly to have length from the list $\{10,50,100,200,500,1000\}$ and generated uniform-randomly from the Pauli gate set.

\begin{figure}
\includegraphics[width=\columnwidth]{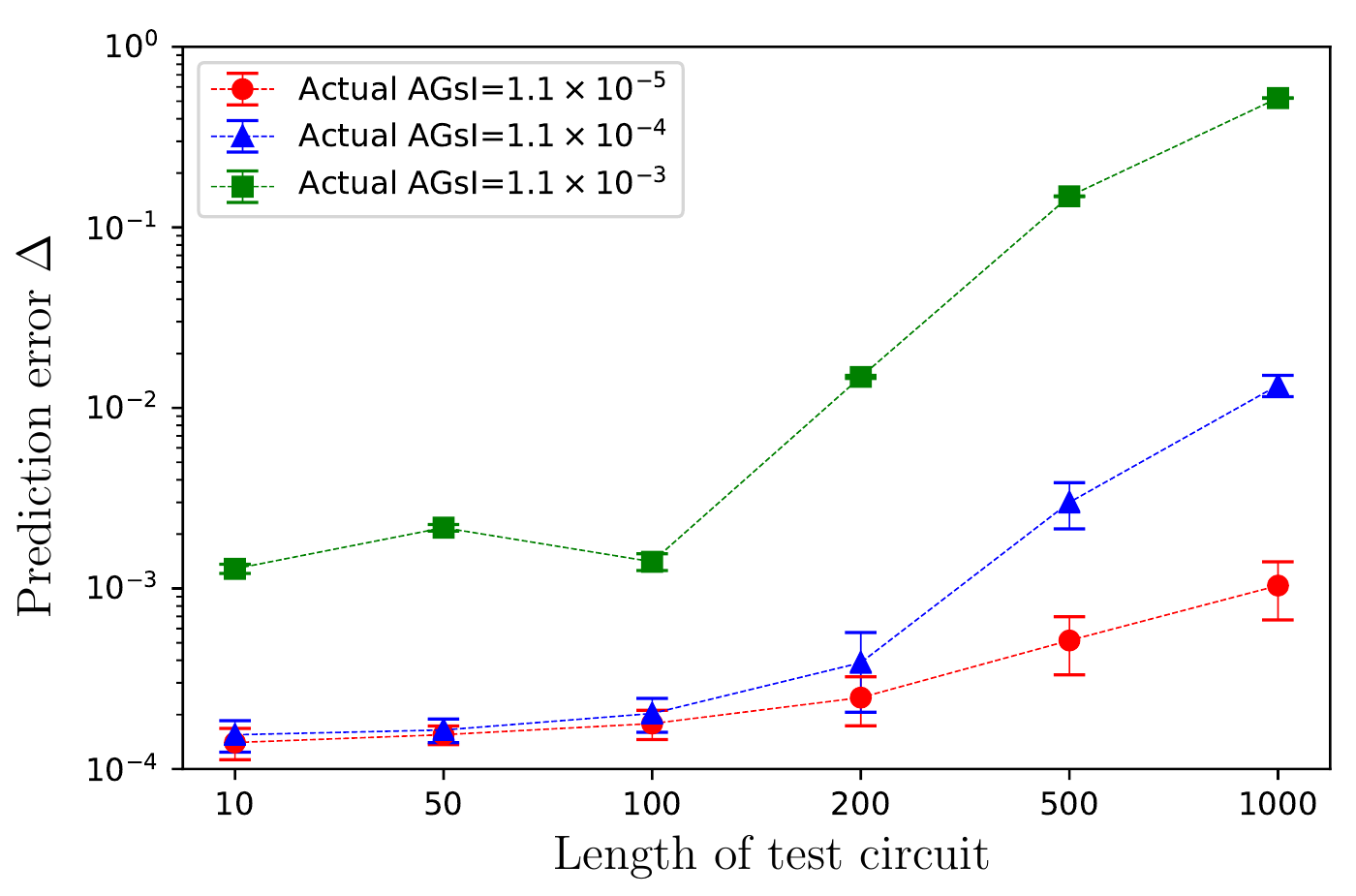}
\caption{\label{fig:perror_xyz} Prediction errors for the incomplete gate set $\{X, Y, Z\}$ using simulated experiments. Three sets of true noise parameters, corresponding to AGsI values of $1.1\times 10^{-5}$, $1.1\times 10^{-4}$, and $1.1\times 10^{-3}$, were used. Tomographic circuits were generated uniform-randomly from the gate set with varying lengths. For each set of noise parameters, the entire simulation was repeated 10 times to generate the error bars.}
\end{figure}

Figure \ref{fig:perror_xyz} shows the prediction error $\Delta$ for three different ``true" noise strengths, as quantified by the \mbox{AGsIs} computed from the true $\ve$ parameters. In all three cases, the prediction errors are small, and they grow as the length of the test circuit grows, to be expected not just because of an accumulation of estimation errors, but also because we approach the boundary of $L\epsilon\sim 1$ for which our linear approximation fails. For this case of incomplete gate set, we show no comparison with standard GST, which does not accommodate such a situation.

\medskip
\noindent\textit{Complete gate set.} As a second example, we consider characterization of tomographically complete gate sets. The first gate set we consider is $\{\id, g_x\equiv X_{\pi/2},g_y\equiv Y_{\pi/2}\}$. Here, we explore different ways of choosing the $L^{(i)}$s for our randomized circuits, and also compare with the GST results. For a fair comparison, we need to fix the total resource cost of each option. We opted for total number of operations---gates, initializations, and readouts---as our measure of resource cost. Since GST is the most restrictive in terms of its requirements on the circuit structure, we use it to set the resource cost. As we will explain below, the GST circuits we run for our simulated example require a total of about 48,150 operations. We take this also as the resource target for the variants of RL-GST explored here.

\begin{figure*}
\includegraphics[width=\textwidth]{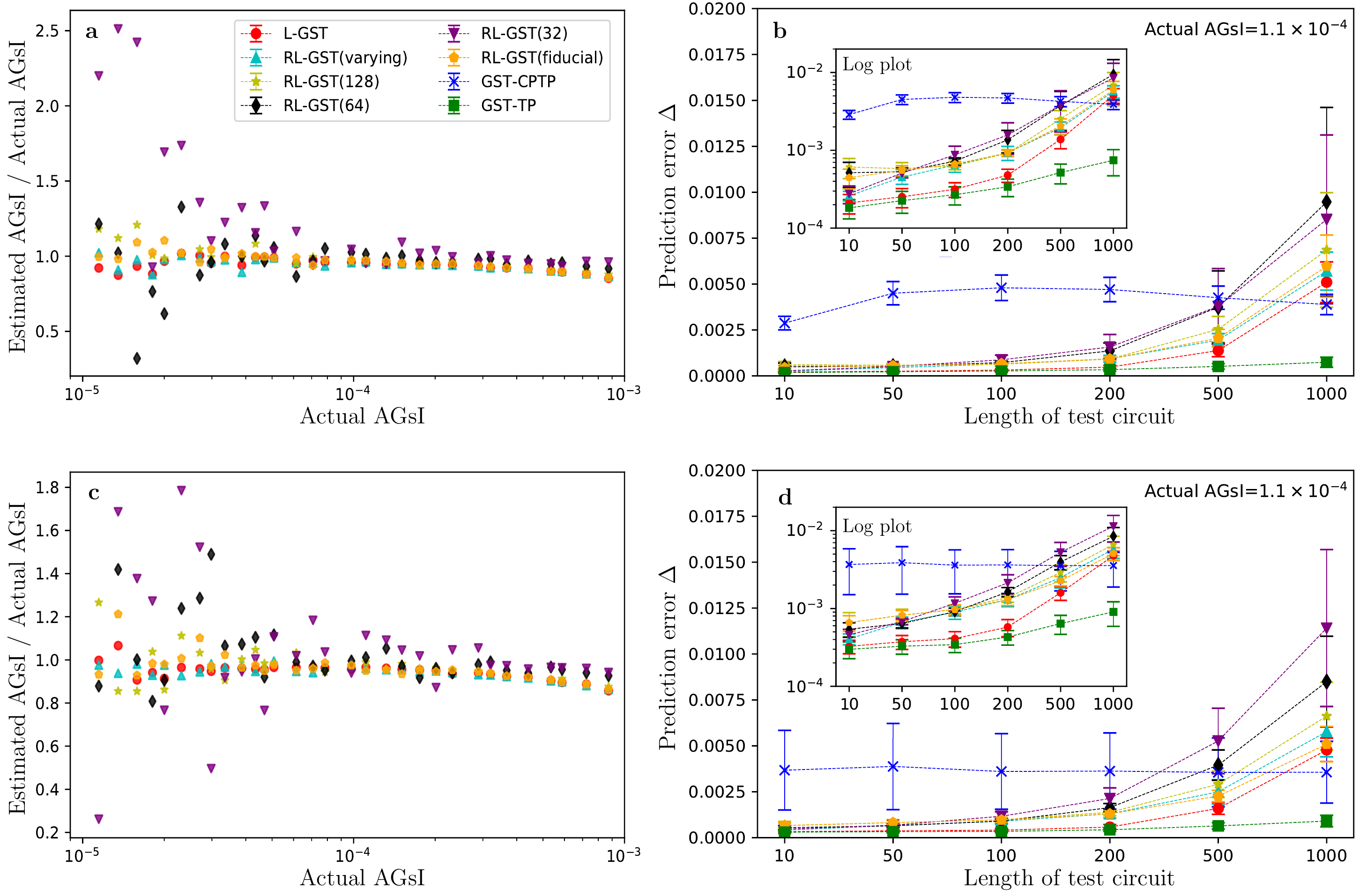}
\caption{\label{fig:comp} Simulated experiments for single-qubit gate sets: plots \textbf{a} and \textbf{b} are for $\{I, X_{\pi/2}, Y_{\pi/2}\}$; plots \textbf{c} and \textbf{d} are for $\{I, T, H\}$. The legend is for all four plots. \textbf{a} and \textbf{c}: The ratios of the estimated AGsI [Eq.~\eqref{agsi}] to the actual AGsI computed from the true parameters are shown for our RL-GST procedure, with different choices of tomographic circuit lengths (32, 64, 128, and varying). In addition, we show the ratios for L-GST, referring to the use of the GST tomographic circuits, and for RL-GST (fiducial), referring to tomographic circuits with the fiducial $F_i,F_j'$ segments but random gates in between. Error bars are generated by repeating simulation for each set of noise parameters 10 times; the sizes of the resulting error bars are similar for all the data. For plot \textbf{a}, the average---over the different sets of noise parameters---error bars are $\{2.0, 2.3, 7.0, 13.1, 25.1, 4.2\}\times 10^{-6}$, listed in the order of appearance in the legend (L-GST and RL-GST variants); for plot \textbf{c}, the average error bars are $\{2.2, 3.1, 7.1, 14.0, 26.5, 4.4\} \times 10^{-6}$. \textbf{b} and \textbf{d}: The prediction errors for the different variants of RL-GST, L-GST, as well as GST procedures are shown, for true noise parameters with actual AGsI $= 1.1\times 10^{-4}$. The error bars are generated by repeating the simulation for each set of noise parameters 10 times. The inset shows the same plot on a logarithmic vertical scale. Note that there are no GST data in plots \textbf{a} and \textbf{c}, as the AGsI is not an invariant quantity for GST.}
\end{figure*}

We examine 4 possibilities for the choices of $L^{(i)}$s in RL-GST: (1) fixed $L^{(i)}=32$; (2) fixed $L^{(i)}=64$; (3) fixed $L^{(i)}=128$; (4) choose $L^{(i)}$ uniform-randomly from the list $\{8,16,32,64,128\}$. To attain the resource cost of about 48,150 operations, we choose 1416, 730,370, and 187 circuits, respectively, for the four options, together with a null circuit for each case. Altogether, these choices correspond to 48,146, 48,182, 48,102, and 48,248 (expected) operations, respectively. The actual number of circuits and operations used in the simulations are often a little fewer, as we discard, without replacement, any duplicate circuits generated in the random drawing.

The GST comparison is done with help of the pygsti Python package. In GST, each tomographic circuit has the structure $F_j' + \textrm{germ}^k + F_i$, where $F_i$ and $F_j'$ are fiducial gate sequences corresponding to tomographically complete state preparation and measurement (the $i,j$ indices label which state and which measurement outcome, respectively, while the ``germ" is some chosen short gate sequence repeated $k$ times to amplify the effect of the noise. Here, we set $\{F_i\}$ and $\{F_j'\}$ to both be the set $\{\textrm{null},g_x,g_y,g_x^2,g_x^3,g_y^3\}$. The set $\{F_i|0\rangle\}$ are then precisely the standard process-tomographic states, $\{|0\rangle,|-\upi\rangle,|+\rangle,|1\rangle,|\upi\rangle,|-\rangle\}$, where $|\pm\rangle\equiv \frac{1}{\sqrt 2}(|0\rangle+|1\rangle)$ and $|\pm\upi\rangle\equiv\frac{1}{\sqrt2}(|0\rangle\pm\upi|1\rangle)$; the set $\{\langle 0|F_j'\}$ are then the adjoints of the standard 6-state measurement (i.e., measure the Pauli $X$, $Y$, $Z$ observables) outcomes. The germs we make use of are those suggested by the GST Python package: $\id, g_x,g_y,g_yg_x$, and $g_yg_xg_x$; each germ is repeated for $k$ values taking values $2^0,2^1,\ldots$, and $2^7$. The maximum value of $2^7$ is chosen to ensure $L\epsilon\ll 1$ for our simulated noise strength, while the powers-of-2 increments are suggested by the Python package. In addition, we also include circuits of the form $F_j'+F_i$ for different $i,j$s, corresponding to taking a ``null" germ. Duplicate circuits are removed. Altogether, we have 1,120 circuits, involving a total of 48,149 operations.
 
As an added angle for exploring the importance of the germ design in GST, we retain the $F_i$ and $F_j'$ state preparation and measurement fiducial sequences in the GST circuits, but replace the middle GST-prescribed $\textrm{germ}^k$ by a uniform-randomly chosen sequence, each of fixed length $L^{(i)}=2^7$, of gates from the gate set. We use 10 different random gate sequences altogether. We also include, as before, circuits with the structure $F_j'+F_i$. Altogether, we have 384 circuits (after eliminating duplicates), with a total of 48,133 operations. 

Linear inversion, as described in our RL-GST procedure earlier, is used to obtain the estimators for all the variants of RL-GST differing in the choice of $\vC$. For the GST circuits, we compute the estimators from the optimization carried out by the GST Python package, with or without the CP option (both estimators are TP). For added comparison, we use our linear inversion procedure for the data obtained from the GST tomographic circuits (labeled L-GST in Fig.~\ref{fig:comp}). The results are given in Figs.~\ref{fig:comp}a and b. Figure~\ref{fig:comp}a plots the AGsI estimates for the gate set; Fig.~\ref{fig:comp}b gives the prediction errors for 1000 test circuits (as in the incomplete gate set example), randomly chosen from the gate set and of lengths $\{10,50,100,200,500,1000\}$. The test circuits in Fig.~\ref{fig:comp}b are for a single randomly chosen set of true noise parameters with AGsI$=1.1\times 10^{-4}$. From the plots, our four RL-GST variants, with different $L^{(i)}$ choices have comparable performance, though the shorter circuits have larger scatter in the AGsI predictions. The one with varying $L^{(i)}$ lengths has the smallest prediction error, but the difference is small, suggesting a robustness against the specific choice of $L^{(i)}$. All four do somewhat worse than GST, but one should note the much lower time cost for nearly the same performance: For each run of the protocol (getting the estimate $\widehat\ve$ from one set of tomographic circuits), RL-GST tomography procedure took under 1 second to complete on a regular laptop. The GST procedure, either the TP or the CPTP options provided by the Python package, took about 30 seconds on the same laptop for the GST-prescribed circuit set. The time-cost advantage for RL-GST becomes even more pronounced with the multiqubit examples below. 

Figures \ref{fig:comp}a and b also offer lessons about GST. Observe that the variant with the $F_i,F_j'$ structure, but now sandwiching a random sequence, does similarly as our RL-GST options, suggesting that the a proper choice of germ is important for GST to work well. Also notable in Fig.~\ref{fig:comp}b is the close performance of the linear inversion estimator to that of the GST-optimized estimator, with prediction errors differing more only for longer circuits, a direct consequence of our linear approximation. As the optimization to reconstruct the estimator is the time-expensive component of GST, this suggests that one could opt for a simple linear inversion while retaining the structure of the GST tomographic circuits.

We repeat the entire simulation experiment for a different single-qubit gate set, $\{\id, H, T\}$, where $H$ is the Hadamard gate $H\equiv|+\rangle\langle 0|+|-\rangle\langle 1|$, and $T\equiv |0\rangle\langle 0|+\upe^{\upi\pi/4}|1\rangle\langle 1|$. In this case, the GST fiducial sequences are chosen as $F_i\in\{I, H, HT^4H, T^2H, T^4H, T^6H\}$ and $F_j'\in\{I, H, HT^4H, HT^6, HT^4, HT^2\}$. Both again generate the 6-state preparation and measurement as used for the $\{\id, g_x,g_y\}$ gate set, but the preparation and measurement sequences are different here since $T^\dagger = T^7\neq T$. The germs, as prescribed by the GST Python package, are $I$, $T$, $H$, $HT^3$, $T I^4$, $T^2HI^3$, $HTHI^3$, $H^4T$, $THT^3I$, and $HTH^2T^2$. As before, we  pick our RL-GST random circuits to match the number of operations used in the GST-prescribed set. The results are given in Figs.~\ref{fig:comp}c and d.

From the two gate sets, we draw the conclusion that RL-GST offers a quick estimate of the noise, providing comparable---though slightly worse---results as GST, but in a much shorter time. One also need not be too bothered with the choice of tomographic circuits as randomly chosen circuit sequences with different $L^{(i)}$ values work nearly as well as the carefully crafted ones from GST.

\bigskip
\noindent\textbf{Two-qubit simulations.}
Next, we simulate a two-qubit system with crosstalk betwen the two qubits. To each two-qubit gate, we attach a noise map of the form $e(\ve)\equiv [\mathrm{AD}(\upe_0)\otimes\mathrm{AD}(\upe_1)]\circ\mathrm{Pauli2}(\upe_2,\upe_3,\ldots,\upe_{16})\circ\{[R_x(\upe_{17})\circ R_y(\upe_{18})\circ R_z(\upe_{19})]\otimes[R_x(\upe_{20})\circ R_y(\upe_{21})\circ R_z(\upe_{22})]\}$. Here, $\mathrm{AD}$, $R_x$, $R_y$ and $R_z$ are as in the single-qubit situation, while $\mathrm{Pauli2}$ is the two-qubit Pauli map---the source of the two-qubit cross-talk---$\mathrm{Pauli2}(\upe_2,\ldots,\upe_{16})(\,\cdot\,)=\sum_{i,j=0,x,y,z}q_{ij}\sigma_i\otimes \sigma_j(\,\cdot\,)\sigma_i\otimes \sigma_j$, with $q_{0x}=\upe_2,q_{0y}=\upe_3,q_{0z}=\upe_4,q_{x0}=\upe_5,q_{xx}=\upe_6$, etc., and $q_{00}$ is determined by $\sum_{i,j=0,x,y,z}q_{ij}=1$. The noisy input state is the product state $\widetilde\rho_\mathrm{in}=\widetilde\rho_\mathrm{in}^{(1)}\otimes \widetilde\rho_\mathrm{in}^{(1)}$ as used in the single-qubit simulation; the noisy $Z$ measurement on each qubit (no crosstalk) is assumed to comprise the outcomes (POVM) operators $\widetilde P_0^{(1)}$ and $\id-\widetilde P_0^{(1)}$, where $\widetilde P_0^{(1)}=\widetilde \rho_\mathrm{in}^{(1)}$.

\begin{figure}
\includegraphics[width=\columnwidth]{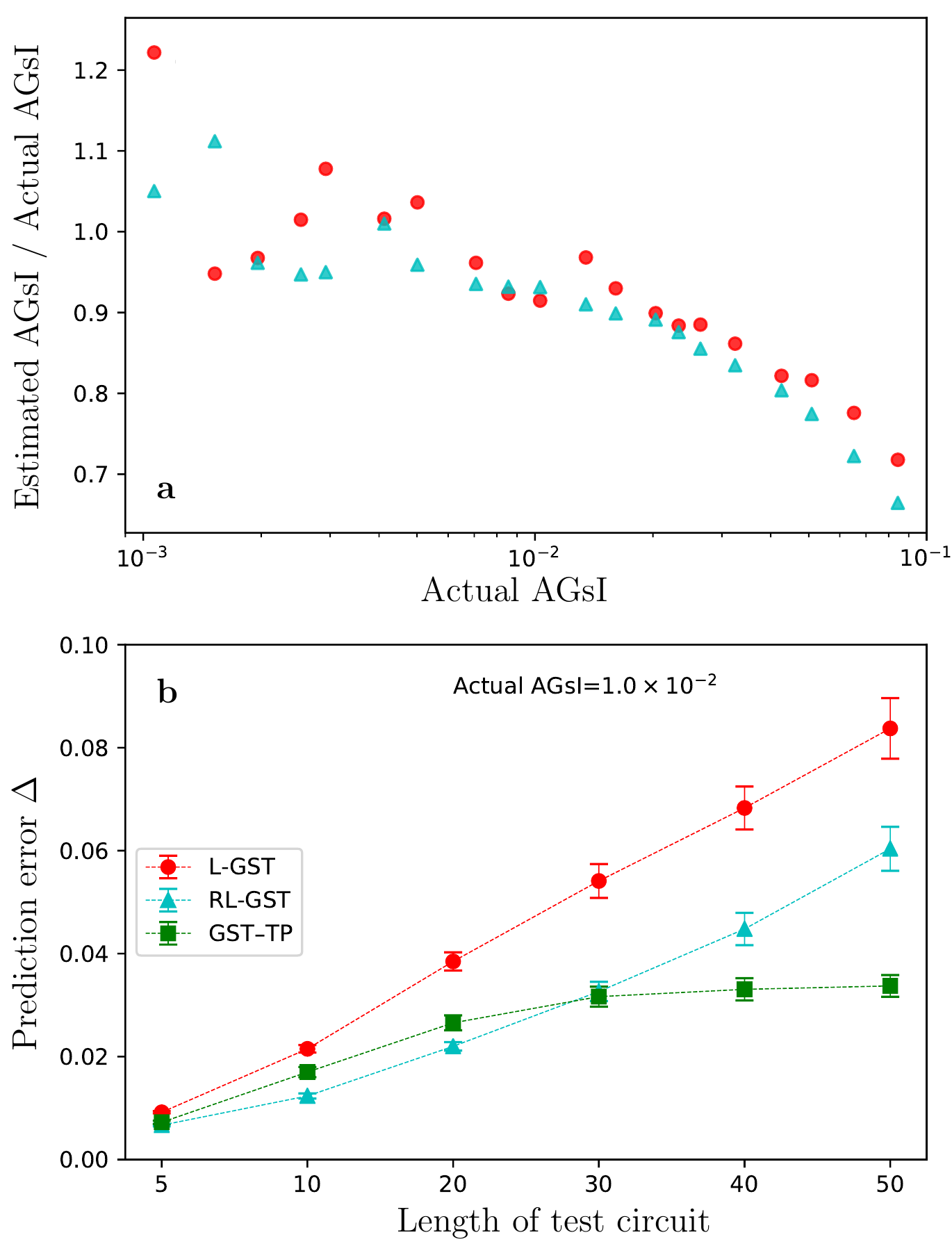}
\caption{\label{fig:ixy2q}Simulation results for a two-qubit gate set. \textbf{a.} 
The ratios of the estimated AGsI [Eq.~\eqref{agsi}] to the actual AGsI computed from the true parameters are shown for the RL-GST and L-GST procedures, with error bars of average sizes $2.6\times 10^{-4}$ and $7.5 \times 10^{-4}$, respectively. \textbf{b.} The prediction errors for RL-GST, L-GST, and GST procedures are shown, for true noise parameters with actual AGsI $= 1.1\times 10^{-2}$. The error bars are generated by repeating the simulation for each set of noise parameters 10 times. The legend is for both plots \textbf{a} and \textbf{b}; there are no GST data in plot \textbf{a}, as the AGsI is not an invariant quantity for GST.
}
\end{figure}

The gates on the two qubits are assumed to be $\id,g_x=X_{\pi/2}, g_y=Y_{\pi/2}$ for each of the qubits, and a two-qubit CNOT gate with one of the qubits (qubit 1) as the control, the other (qubit 2) as the target. With the expectation of crosstalk between the qubits, the gate set we are to characterize comprises all possible two-qubit combinations of the gates, namely, the $\CNOT$ together with $\{g_{ij}\}_{i,j=0,x,y}$, where $g_{ij}\equiv g_i\otimes g_j$ with $i,j=0,x,y$ and $g_0\equiv \id$.

As in the single-qubit simulations, we impose a common resource cost, set by the GST procedure. Because running the full GST algorithm for the two-qubit situation turns out to be very time-consuming, we use simpler circuits here. 
The fiducial sequences $\{F_i\}$ and $\{F_j'\}$ are taken from the set $\{\textrm{null},g_{0x}, g_{0y},g_{0x}g_{0x}, g_{x0}, g_{xx}, g_{xy}, g_{0x}g_{xx}, g_{y0}, g_{yx},g_{yy},$ $g_{0x}g_{yx}, g_{x0}g_{x0}, g_{x0}g_{xx}, g_{x0}g_{xy}, g_{xx}g_{xx}\}$. Ideal versions of the fiducial sequences map the input $|0\rangle$ state to the 16 states $\{|0\rangle,|-\upi\rangle,|+\rangle,|1\rangle\}^{\otimes 2}$. Furthermore, we choose the germs for the GST circuit as individual gates in the gate set, and each germ is applied only once ($k=1$). There are 2,336 GST-type circuits generated this way, with a total of 13,612 operations. For our RL-GST procedure (no $F_i$--$F_j'$ structure), we choose ${\left\lceil\frac{13612}{8+2}\right\rceil}=1,362$ circuits, with 8 gates in each circuit; the null sequence is again added in as an additional circuit.

Figure \ref{fig:ixy2q} shows the results for the two-qubit situation. For GST, for shorter runtime, we ran only the case with TP (not necessarily CP) constraints, more comparable to our linear-inversion estimator. One single GST run took about 35 minutes to complete, versus just 12 seconds for RL-GST, which achieved similar estimation accuracy as is apparent in Fig.~\ref{fig:ixy2q}. Again, we used the data from the GST circuits to compute the linear-inversion estimator (L-GST); the results are also given in Fig.~\ref{fig:ixy2q}, this time doing more poorly than our RL-GST and the original GST estimator.

\bigskip
\noindent\textbf{Characterization of IBM quantum devices.}
Next, we apply our method to characterizing the noise in three of the 5-qubit devices publicly available on the IBM Quantum Experience platform \cite{ibmQX}. The IBM devices make use of primitive gates $g_x=X_{\pi/2}$ and $g_y=Y_{\pi/2}$ to accomplish a variety of Clifford gate operations. We hence consider a gate set comprising the single-qubit gates $\{\id, g_x,g_y\}$ for each of the 5 qubits in the device, together with the two-qubit $\CNOT$ gates specific to each device. The default (ideal) input state is $|0\rangle$ for each qubit; $Z$ is measured on each qubit at the end of every circuit.

\begin{figure*}
\includegraphics[width=\textwidth]{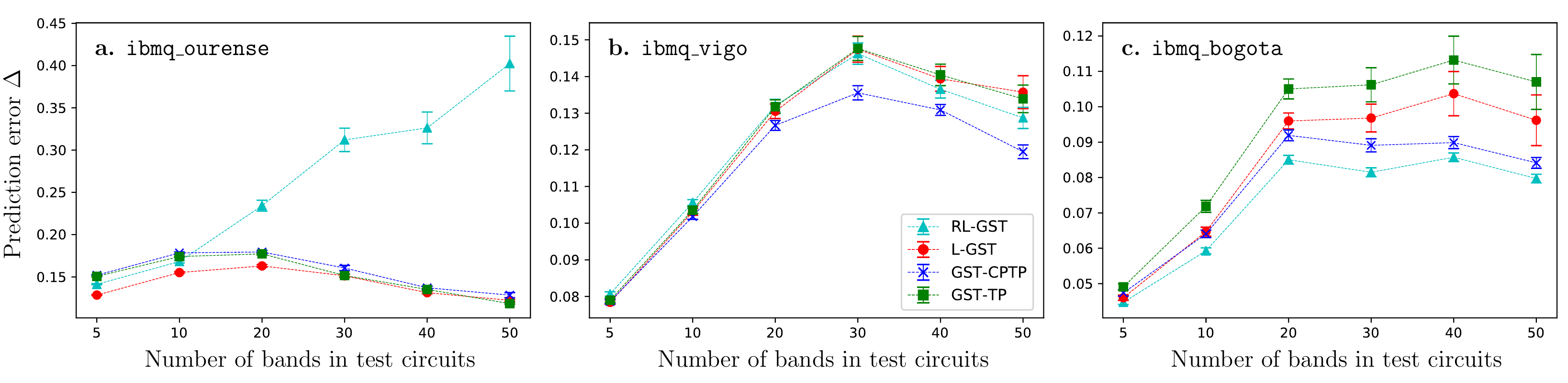}
\caption{\label{fig:IBMperr} Prediction errors of 5-qubit test circuits on three IBM devices for the different estimation procedures. All experiments were done on 10 September 2020. The error bars indicate our uncertainty in the estimates $\widehat\ve$, generated by bootstrapping (100 samples) the tomographic experimental data. For \vigo~and \bogota, all estimation procedures give similar prediction errors, with RL-GST taking the least computational time and design effort. For \ourense, all methods show worse prediction errors, with RL-GST performing more poorly than other procedures, a likely reason being the time-drift in the noise of the device. The goodness-of-fit quantities (see Table \ref{tab:12qubit} below) for \ourense~are large for all procedures, so the model fit is poor anyway, and one cannot draw reliable conclusions about the performance of the schemes; see main text for further discussion.}
\end{figure*}

For a 5-qubit device, the most general approach would be to assign a 5-qubit noise map for every single gate. For example, the ideal single-qubit $g_x$ gate applied on qubit 0 should be thought of as $g_x\otimes\id\otimes\id\otimes\id\otimes\id$ with a 5-qubit noise map to be characterized. In principle, our RL-GST approach can accommodate such a scenario, as can the GST procedure. The sheer combinatorial variety of gates, and hence the number of noise parameters, that one would have to handle, however, are forbidding, and likely ultimately not useful.

Instead, we opt for a more practical route: First characterize the noise in the single-qubit gates individually, as if they are done on separate single qubits, and then use these single-qubit gates, now with known (estimated) noise, to characterize the noise in the two-qubit gates. This assumes that the crosstalk between qubits in the device is unimportant, which may or may not hold for the device in question. We discuss goodness-of-fit measures at the end to gauge the suitability of our assumed model.

We characterize the noise in three IBM 5-qubit devices: \ourense~\cite{ourense}, \vigo~\cite{vigo}, and \bogota~\cite{bogota}. To characterize the single-qubit gates, we make use of both our randomized tomographic circuits as well as the GST-type circuits. 
For the GST circuits, we follow the procedure as in our single-qubit simulation for the same $\{\id,g_x,g_y\}$ gate set described earlier, having the structure $F_j'+\textrm{germ}^k+F_i$, and the fiducial sequences are chosen as before (the 6-state tomographic states and measurement). Because of the time-drift in the IBM devices, we found it necessary to use tomographic circuits that are not too long. Consequently, the $k$ values are restricted to $k=1,2$ and $4$ only. This gives 285 tomographic circuits $\{C^{(i)}\}$. For RL-GST, to maintain the same resource cost as for GST, we randomly choose 83 circuits each for circuit lengths 8 and 16; a null circuit is again included, as before. The same tomographic circuit is run on each of the 5 qubits of the device, and submitted together to the IBM platform, to be run simultaneously. The number of shots per circuit is $N_s=8192$. 

Tomographic circuits for the two-qubit gates are constructed simply as $F_j'+\CNOT+F_i$, where the $F_j'$ and $F_i$ are now the fiducial sequences used in our two-qubit simulations above. When doing the reconstruction of the noise parameters for the two-qubit gates, the noise parameters for the single-qubit are set to be those estimated from the single-qubit characterization rounds. The two-qubit gate characterization thus looks exactly like what one would do in standard process tomography. For simplicity, we use linear inversion (respecting the TP, but not CP, constraint) to obtain an estimate of the two-qubit noise process.

Figure \ref{fig:IBMperr} shows the prediction errors for the three different devices, using our RL-GST as well as the GST procedures (TP and CPTP variants) for the single-qubit gate characterizations; we include also the data for L-GST, where our linear-regime analysis is applied to GST tomographic circuits. The test circuits are 5-qubit circuits. Each test circuit is composed of multiple time-bands. Each time-band is randomly chosen (50-50) to comprise either only single-qubit gates or single- and two-qubit gates. In a single-qubit-gate time-band, a single-qubit gate is chosen uniform-randomly from the gate set $\{\id,g_x,g_y\}$ for each qubit. In a time-band chosen to have two-qubit gates, we first uniform-randomly pick one of the $\CNOT$ gates to apply on the appropriate control and target qubits; the other qubits are each assigned a sequence of randomly chosen single-qubit gates, of a length that occupies the time taken for the $\CNOT$ gate in the same time-band to be completed. The prediction errors are computed as in Eq.~\eqref{eq:statDist}, with $p_\mu$s set as the relative frequencies obtained from the IBM device itself. 

Now, the IBM devices are well known to suffer from time-drifts in the gate noise, a situation not accommodated within our inference model. Crosstalk between qubits is another source of possible model deviation. Given these possible deviations to our assumed model, it is prudent to perform a goodness-of-fit test, to assess the reliability of our estimates for the single-qubit gates. A common approach is to use the $\chi^2$ statistic, defined in this case as
\begin{equation}
    \chi^2=\sum_{i=1}^{N_c}\Bigl(\frac{p_i-\widehat{p}_i}{\Sigma_i}\Bigr)^2.
\end{equation}
Here, $N_{\mathrm{c}}$ is the number of circuits, $p_i$ and $\widehat{p}_i$, as before, are the measured and predicted (using $\widehat\ve$) outcome probabilities of noise parameters, and $\Sigma_{i}$ is the standard deviation of the measured $p_i$s, approximated as $\Sigma_{i}=\Bigl[\frac{p_i(1-p_i)}{N_{\mathrm{s}}}\Bigr]^{1/2}$. The quantity $\chi^2$ satisfies a $\chi^2$-distribution with $f\equiv N_{\mathrm{c}}-N_{\mathrm{p}}$ degrees of freedom, where $N_{\mathrm{p}}$ is the number of free (error) parameters. Following Ref.~\cite{blume2017demonstration}, we quantify goodness of fit by $n_{\sigma}$, defined as
\begin{equation}\label{eq:sigma}
    n_{\sigma}\equiv\frac{\chi^2-f}{\sqrt{2f}}\,.
\end{equation}

Table~\ref{tab:12qubit} gives the data for the single- and two-qubit gates for the IBM devices: the estimated AGsI values from the RL-GST, L-GST, and GST procedures, and the single-qubit goodness-of-fit measure $n_\sigma$. Comparing the goodness-of-fit numbers with the performance shown in Fig.~\ref{fig:IBMperr}, we see that prediction errors become large when the goodness of fit numbers exceed $\sim100$. The fit is mostly good for \vigo~and \bogota, and this is reflected in the lower prediction errors in Figs.~\ref{fig:IBMperr}b and c, while the higher prediction errors for \ourense~is mirrored by the large $n_\sigma$ values in Table~\ref{tab:12qubit}. We suggest not proceeding with the two-qubit gate characterization unless the single-qubit $n_\sigma$ values are sufficiently small. 

The prediction errors for \vigo~and \bogota~are comparable across the RL-GST, L-GST, and GST procedures, with RL-GST taking the least computational time and design effort. RL-GST seems more susceptible to the issue of poor fit, as apparent in the plot for \ourense, but this could also be due to the fact that the tomographic data for the randomized circuits used in RL-GST were taken earlier in time, and hence with a longer time-gap to the test-circuit runs, than the GST circuits. The noise in the test circuits may have drifted further from that captured in the RL-GST data than in the GST data. Another notable feature in Fig.~\ref{fig:IBMperr} is the slight decrease in prediction errors as the number of bands ($>20$) in the test circuits increases. We suspect this is due to the state of the quantum computer stabilizing towards an equilibrium value, regardless of the circuit applied, as observed empirically from the output measurement statistics, as the number of gate operations increases.

\begin{table*}
\caption{\label{tab:12qubit}%
Data for single- and two-qubit gates in the IBM
devices (\ourense, \vigo, and \bogota). All experiments were performed on 10
September 2020 and the IBM AGsI values were from the calibration cycle
closest to the experiment date. For single-qubit gates; The estimated AGsI
values are computed from the RL-GST, L-GST, and GST (CPTP and TP
constraints) procedures. Negative infidelity values can arise because we do
not impose CP constraints in the estimation. Each single-pulse single-qubit
gate takes about 36ns to complete. For two-qubit gates: 
$\mathrm{cx}a\_b$ refers to a $\CNOT$ gate between qubits $a$ and $b$, with
qubit $a$ as the control, and qubit $b$ as the target. The RL-GST, L-GST,
GST-CPTP, and GST-TP headings refer to the procedure used for estimating the
noise parameters for the \emph{single}-qubit gates. In every case, the
two-qubit gate noise parameters are estimated using standard linear
inversion. Negative infidelity values for the two-qubit gates can thus arise. } 

\begin{tabular}{c|cccd@{\quad}r@{$\,\pm\,$}lr@{$\,\pm\,$}lcr@{$\,\pm\,$}lr@{$\,\pm\,$}lr@{$\,\pm\,$}lr@{$\,\pm\,$}l}
\hline\hline
  \multirow{17}{0.4cm}{%
  \rotatebox{90}{\hspace*{-1cm}\textbf{single-qubit gates}}}%
  &&Qubit &\quad \qquad&\multicolumn{5}{c}{AGsI ($\times 10^{-4}$)}%
  &\qquad \qquad&\multicolumn{8}{c}{Goodness of fit $n_\sigma$}\\
  &&&&\multicolumn{1}{l}{IBM}& \multicolumn{2}{c}{RL-GST}& \multicolumn{2}{c}{L-GST}  &&%
  \multicolumn{2}{c}{RL-GST}& \multicolumn{2}{c}{L-GST}& \multicolumn{2}{c}{GST-CPTP}&\multicolumn{2}{c}{GST-TP}  \\
\cline{2-18}
&\multirow{5}{*}{\rotatebox{90}{\hspace*{-0.3cm}\ourense}}
&$\mathrm{Q}_0$ &&7.0 &$14.3$&$0.9$  &$18.3$&$1.3$  && $23$&$3$ &$30$&$3$ &$30$&$4$ &$28$&$3$ \\
&&  $\mathrm{Q}_1$ &&4.1 &$13.1$&$0.9$ &$22.1$&$1.2$ &&$32$&$3$ &$35$&$3$ &$308$&$63$ &$24$&$3$    \\
&& $\mathrm{Q}_2$ &&1.7 &$2.9$&$0.6$  &$-0.03$&$1.0$  &&$6$&$ 2$ &$4$&$2$& $13$&$ 20$ &$2$&$2$ \\  
&& $\mathrm{Q}_3$ &&4.5 &$8.4$&$1.2$ &$126.3$&$2.3$ &&$1040$&$ 20$ &$1582$&$18$ &$1630$&$41$ &$1215$&$15$ \\ 
&& $\mathrm{Q}_4$ &&10.8  &$20.8$&$1.4$ &$59.8$&$2.4$  &&$ 843$&$14 $&$1885$&$22$ & $1961$&$128$ &$1753$&$21$ \\[1ex]
\cline{2-18}
&\multirow{5}{*}{\rotatebox{90}{\hspace*{-0.3cm}\vigo}} 
&$\mathrm{Q}_0$ &&3.5 &$1.5$&$0.6$ &$5.5$&$1.1$ &&$46$&$ 7$ &$22$&$4$ &$17$&$4$ &$14$&$3$ \\  
&&$\mathrm{Q}_1$ &&4.8 &$12.1$&$0.7$ &$11.9$&$1.0$ && $78$&$ 6$ &$66$&$6$ &$47$&$4$ &$43$&$4$   \\  
&&$\mathrm{Q}_2$ &&3.9 &$7.0$&$0.6$ &$4.5$&$1.1$ && $28$&$4$ &$25$&$4$ &$21$&$3$ &$18$&$2$\\  
&&$\mathrm{Q}_3$ &&4.8 &$7.6$&$0.7$ &$4.3$&$1.0$ && $45$&$5$ &$44$&$3$ &$45$&$7$ &$41$&$3$ \\ 
&&$\mathrm{Q}_4$ &&7.2  &$7.4$&$0.5$ &$6.6$&$0.9$ && $30$&$4$ &$30$&$4$ &$33$&$5$ &$28$&$3$ \\[1ex]
\cline{2-18}
& \multirow{5}{*}{\rotatebox{90}{\hspace*{-0.3cm}\bogota}} 
&$\mathrm{Q}_0$ &&2.8 &$3.3$&$0.6$ &$5.6$&$1.1$ &&$4$&$2$ &$7$&$2$ &$13$&$3$ &$6$&$2$ \\  
&&$\mathrm{Q}_1$ &&1.9 &$7.3$&$0.6$ &$6.6$&$1.1$ && $28$&$4$ &$13$&$3$ &$13$&$9$ &$11$&$2$   \\  
&&$\mathrm{Q}_2$ &&1.8 &$3.8$&$0.8$ &$-1.1$&$1.2$ && $6$&$2$ &$527$&$16$ &$524$&$16$ &$508$&$13$\\  
&&$\mathrm{Q}_3$ &&2.9 &$6.5$&$0.7$ &$5.8$&$1.0$ && $7$&$3$ &$11$&$3$ &$26$&$5$ &$6$&$2$ \\ 
&&$\mathrm{Q}_4$ &&2.6  &$5.3$&$0.8$ &$5.2$&$1.3$ && $9$&$3$ &$5$&$3$ &$10$&$3$ &$2$&$2$ \\[1ex]
\hline\hline
\end{tabular}

\bigskip

\begin{tabular}{c|cccccc r@{$\,\pm\,$}l r@{$\,\pm\,$}l r@{$\,\pm\,$}l r@{$\,\pm\,$}l}
\hline\hline
&&Gates    &\quad& Gate time&\quad\quad&&\multicolumn{8}{c}{Average gate infidelity ($\times 10^{-2}$)} \\
  &&&&(ns)&&IBM & \multicolumn{2}{c}{RL-GST}
  &\multicolumn{2}{c}{L-GST}& \multicolumn{2}{c}{GST-CPTP} & \multicolumn{2}{c}{GST-TP} \\
\cline{2-15}
\multirow{26}{0.4cm}{\rotatebox{90}{\hspace*{-0.3cm}\textbf{two-qubit gates}}}&\multirow{8}{*}{\rotatebox{90}{\hspace*{-0.4cm}\ourense}}  
&$\mathrm{cx}0\_1$ &&270.2 &&   1.11  &$-1.53$&$0.20$    &$-1.14$&$0.15$  &$-1.75$&$0.17$ &$-0.98$&$0.16$ \\
&&$\mathrm{cx}1\_0$ &&234.7 &&   1.11  &$-0.68$&$0.20$    &$-0.29$&$0.16$  &$-1.00$&$0.17$ &$-0.39$&$0.16$\\
&&$\mathrm{cx}1\_2$ &&391.1 &&   0.74  &$-0.38$&$0.21$    &$-0.48$&$0.16$  &$-1.21$&$0.23$ &$-0.47$&$0.17$\\
&&$\mathrm{cx}2\_1$ &&426.7 &&   0.74  &$0.07$&$0.21$    &$-0.03$&$0.14$  &$-0.56$&$0.23$ &$0.23$&$0.14$\\
&&$\mathrm{cx}1\_3$ &&611.6 &&   1.28  &$-11.78$&$0.25$    &$-8.70$&$0.18$ &$-12.23$&$0.28$ &$-9.87$&$0.21$\\
&&$\mathrm{cx}3\_1$ &&576.0 &&   1.28  &$-11.95$&$0.26$    &$-8.87$&$0.19$ &$-12.68$&$0.31$ &$-9.87$&$0.23$\\
&&$\mathrm{cx}3\_4$ &&305.8 &&   0.92  &$-8.71$&$0.27$   &$-5.24$&$0.20$  &$0.60$&$0.71$ &$-0.37$&$0.32$\\
&&$\mathrm{cx}4\_3$ &&270.2 &&   0.92  &$-10.58$&$0.29$    &$-7.10$&$0.21$ &$-1.52$&$0.97$ &$-2.53$&$0.35$\\
\cline{2-15}
&\multirow{8}{*}{\rotatebox{90}{\hspace*{-0.5cm}\vigo}}
&$\mathrm{cx}0\_1$ &&519.1 &&   1.13  &$1.64$&$0.19$    &$1.38$&$0.16$  &$1.82$&$0.17$ &$1.56$&$0.18$ \\
&&$\mathrm{cx}1\_0$ &&554.7 &&   1.13  &$1.23$&$0.16$    &$0.97$&$0.15$  &$1.38$&$0.16$ &$1.13$&$0.16$\\
&&$\mathrm{cx}1\_2$ &&227.6 &&   0.68  &$3.73$&$0.21$    &$2.92$&$0.17$  &$3.35$&$0.17$ &$3.25$&$0.18$\\
&&$\mathrm{cx}2\_1$ &&263.1 &&   0.68  &$3.89$&$0.19$    &$3.08$&$0.17$  &$3.45$&$0.17$ &$3.32$&$0.17$\\
&&$\mathrm{cx}1\_3$ &&497.8 &&   1.33  &$0.58$&$0.19$    &$0.13$&$0.16$ &$0.52$&$0.19$ &$0.21$&$0.17$\\
&&$\mathrm{cx}3\_1$ &&462.2 &&   1.33  &$0.88$&$0.19$    &$0.43$&$0.16$ &$0.82$&$0.20$ &$0.55$&$0.18$\\
&&$\mathrm{cx}3\_4$ &&270.2 &&   0.78  &$0.97$&$0.19$   &$0.87$&$0.15$  &$1.26$&$0.17$ &$0.98$&$0.16$\\
&&$\mathrm{cx}4\_3$ &&305.8 &&   0.78  &$-0.32$&$0.17$    &$-0.42$&$0.15$ &$-0.22$&$0.18$ &$-0.47$&$0.16$\\
\cline{2-15}
 &\multirow{8}{*}{\rotatebox{90}{\hspace*{-0.5cm}\bogota}}
 &$\mathrm{cx}0\_1$ &&433.8 &&   1.02  &$3.17$&$0.17$    &$2.65$&$0.15$  &$3.26$&$0.19$ &$3.10$&$0.16$ \\
&&$\mathrm{cx}1\_0$ &&398.2 &&   1.02  &$2.23$&$0.18$    &$1.71$&$0.16$  &$2.16$&$0.19$ &$1.97$&$0.17$\\
&&$\mathrm{cx}1\_2$ &&376.9 &&   0.58  &$2.73$&$0.22$    &$0.27$&$0.18$  &$1.10$&$0.23$ &$0.29$&$0.21$\\
&&$\mathrm{cx}2\_1$ &&412.4 &&   0.58  &$1.02$&$0.21$    &$-1.44$&$0.16$  &$-0.75$&$0.23$ &$-1.58$&$0.18$\\
&&$\mathrm{cx}2\_3$ &&625.8 &&   0.91  &$1.84$&$0.21$    &$-0.12$&$0.15$ &$0.99$&$0.20$ &$-0.07$&$0.17$\\
&&$\mathrm{cx}3\_2$ &&590.2 &&   0.91  &$2.10$&$0.20$    &$0.14$&$0.15$ &$1.20$&$0.19$ &$0.17$&$0.17$\\
&&$\mathrm{cx}3\_4$ &&369.8 &&   0.96  &$1.16$&$0.19$   &$0.69$&$0.16$  &$1.18$&$0.18$ &$0.90$&$0.17$\\
&&$\mathrm{cx}4\_3$ &&334.2 &&   0.96  &$1.16$&$0.20$    &$0.69$&$0.16$ &$1.20$&$0.18$ &$1.00$&$0.17$\\
\hline\hline
\end{tabular}
\end{table*}

\bigskip
\bigskip
\noindent\textbf{CONCLUSIONS}\\[1ex]
In this work, we introduced the procedure of RL-GST, combining the idea of SPAM-error-free characterization of standard GST with the design ease brought about by our randomized circuit choice and computational speed from our linear-noise approximation. This gave us an easy-to-implement gate set tomography approach that can be done on the fly, without complicated tomographic circuit design or long computational wait-time, but which yielded similar performance as standard GST, as evidenced by our illustrative examples. The assumption of the linear-noise regime is a very reasonable approximation for devices potentially capable of achieving useful quantum information processing. We showed how our procedure can be used to characterize the real quantum computing devices offered by the IBM Quantum Experience Platform, and the estimated noise parameters can be used as inputs via the standard GST optimization approach, if further refinement is needed.

Looking ahead, one can imagine examining how a time-drift model can be accommodated within the linear-noise regime, for a better description of the actual noise in real devices, overcoming the issue of poor fit observed in the \ourense~experiment. Cross-talk between the qubits can already be accounted for within our current model, but more work can be done to reduce the computational load brought about by the exponential growth in the number of noise parameters. An interesting direction currently under investigation within our group is to put in noise models built from elementary noise channels, each with a small number of noise parameters, but which together can describe a large variety of noise processes seen in real devices. We will report the results in due course.

\bigskip
\bigskip
\noindent\textbf{ACKNOWLEDGMENTS}\\
This work is supported by the Ministry of Education, Singapore (through grant number MOE2016-T2-1-130). HKN also acknowledges support by a Centre for Quantum Technologies (CQT) Fellowship. CQT is a Research Centre of Excellence funded by the Ministry of Education and the National Research Foundation of Singapore.
We acknowledge the use of the IBM Quantum Experience Platform for this work. The views expressed are those of the authors and do not reflect the official policy or position of IBM or the IBM Quantum team. We thank My Duy Hoang Long for useful discussions throughout the project.

\bigskip
\noindent\textbf{AUTHOR CONTRIBUTIONS}\\
Y. Gu performed the numerical simulations and the IBM experiments.
All authors contributed to the formulation of the problem, the discussion of the results, and the writing of the manuscript.

\newpage
\noindent \textbf{SUPPLEMENTARY INFORMATION}\\[1ex]
\noindent \textbf{Dependence relations.} Here, we provide further technical details on dependence relations, and in particular, discuss the class of gauge transformations expressed by Eq.~\eqref{eq:gaugeTransf}. A dependence relation between the columns of the design matrix $\vC$ can be expressed as the existence of a non-zero vector $\valpha$ such that
\begin{equation}
\valpha \cdot \vC_\ell =0,
\end{equation}
for any row $\vC_\ell$ of $\vC$.
Any vector in the null space of $\vC$ gives such a relation, and the dimension of the null space is equal to the number of independent dependence relations. 

The transformation in Eq.~\eqref{eq:gaugeTransf} gives a class of dependence relations. We ask how many independent dependence relations can be generated by such a transformation, as determined by a map $\cQ$. We first observe that the first row of $\cQ$ is a zero vector for an informationally complete gate set. In particular, because of the trace-preserving constraints of each unitary map $g_\gamma$, the first row of $\cQ-g_\gamma \cQ g_{\gamma}^{\dagger}$ should be zero vector, that is,
\begin{align}
    0&=\cQ_{0,j}-\sum_{m,k}g_{\gamma;0,m}\cQ_{m,k} g^{\dagger}_{\gamma;k,j}\nonumber \\
    &=\cQ_{0,j}-\sum_{k}\cQ_{0,k}g^{\dagger}_{\gamma;k,j} \nonumber \\
    &= \bigl(\cQ-\cQ g_{\gamma}^{\dagger} \bigr)_{0,j}=\sum_{k}\cQ_{0,k}\bigl(\map{I}-g_{\gamma}^{\dagger}\bigr)_{k,j}
\end{align}
holds for any $\gamma$ and $j$.
The second line results from the fact that $g_{\gamma}$ is a unitary map whose first row is the vector $(1,0,0,\cdots)$ when written in the Pauli operator basis. We get a system of linear equations,
\begin{equation}
    \bigl(\cI-g_{\gamma}^{\dagger}\bigr)^{\mathrm{T}} |\cQ_0\rrangle = 0,
\end{equation}
where $|\cQ_0\rrangle^{\mathrm{T}}$ is the first row of $\cQ$. All maps and vectors are now thought of as written in the Pauli operator basis. Because $g_\gamma$ is a real matrix, we have,
\begin{equation}\label{eq:eigv1}
    g_{\gamma}|\cQ_0\rrangle = |\cQ_0\rrangle.
\end{equation}
Thus, $|\cQ_0\rrangle$ is a simultaneous eigenvector with eigenvalue 1 for all the gates $g_\gamma$. 

For an informationally complete gate set, we show that $|\cQ_0\rrangle$ can only be $c|I\rrangle=(c,0,0,0,\cdots)$ where $c$ is some complex number. We assume that the gate set is a generating set for a group of unitary maps. The space spanned by $\dket{I}$ is an invariant subspace of the group. If there is another $\dket{\map{Q}_0}$ that is orthogonal to $\dket{I}$ and satisfies \Eq{eigv1}, then $\dket{\map{Q}_0}$ spans another invariant space. Because finite-dimensional unitary representations of any group are completely reducible, the complement space of $\dket{I}\dbra{I}$ and $\dket{\map{Q}_0}\dbra{\map{Q}_0}$ is also an invariant subspace whose projector is 
denoted as $\overline\cQ$. The input state $\rhoin$ can be written as $\dket{\rhoin}=\frac{1}{\sqrt{d}}\dket{I}+\dket{\map{Q}_0}\dbra{\map{Q}_0}\rhoin\rrangle+\overline\cQ\dket{\rhoin}$. The states that can be prepared by the gate set thus have the form $g_\gamma\dket{\rhoin}=\frac{1}{\sqrt{d}}\dket{I}+\dket{\map{Q}_0}\dbra{\map{Q}_0}\rhoin\rrangle+\overline{\cQ}\dket{\rho_{\gamma}}$ where $\rho_{\gamma}$ can be any state. However, this set of states cannot form an operator basis, as an operator $\frac{1}{\sqrt{d}}\dket{I}-\dket{\map{Q}_0}\dbra{\map{Q}_0}\rhoin\rangle\rangle$ cannot be written as a linear combination of this set of states. The gate set must hence be informationally incomplete, violating our initial assumption.

Because the input state $\rho_{\mathrm{in}}$ has trace 1, the first element of $\cQ |\rhoin\rrangle$ is zero. Then we get
\begin{equation}
    \dbra{\cQ_0}\rhoin\rrangle =\frac{1}{\sqrt{d}}c=0 \quad \Rightarrow \quad c=0\, .
\end{equation}
Thus the first row of $\map{Q}$ is a zero vector.
Then $\cQ$ can be considered as a vector in the space spanned by $|\sigma_{\va}\rrangle \llangle\sigma_{\va'}|$
where $\va\neq \boldsymbol{0}$, which is $(d^2-1)d^2$-dimensional in size.

It is easy to see that the transformation in Eq.~\eqref{eq:gaugeTransf}, from a $\cQ$ to a $\valpha$, is a a linear map. We can prove that the kernel of this linear map is only $\cQ= 0$. Consider a $\cQ$ in the kernel, then we get $\cQ -g_{\gamma}\cQ g_{\gamma}^{\dagger}=0$ for any $g_{\gamma}$ and $\cQ |\rhoin\rrangle=0$. Thus $\cQ g_{\gamma}|\rhoin\rrangle = g_{\gamma}\cQ|\rhoin\rrangle =0$ holds for any $g_{\gamma}$. Because $g_{\gamma}|\rhoin\rrangle$ form a basis for the operator space, then $\cQ$ can only be zero map. Thus the dimension of the null space of $\vC$ due to the transformation 
in Eq.~\eqref{eq:gaugeTransf} is equal to the dimension of the space of all the $\cQ$, i.e., equal to $d^2(d^2-1)$. 

By vectorizing the map $\map{Q}$, the linear map transforming $\map{Q}$ to $\valpha$ can be represented as  a matrix.  Thus for a given $\valpha$, one can determine whether this $\valpha$ is caused by the transformation by solving a system of linear equations. If there is a solution and the first row of the solution is a zero vector, then the $\valpha$ is generated by Eq.~\eqref{eq:gaugeTransf}. We find all the dependence relations $\valpha$ in our examples are from the transformation in Eq.~\eqref{eq:gaugeTransf}. We conjecture this is true for all informationally complete gate sets but cannot prove this in full at the current juncture.

\end{document}